\documentclass[intlimits,twoside,a4paper]{article}

\usepackage[cp1251]{inputenc}
\usepackage{xcolor}
\usepackage[eqsecnum]{cmpj3}

\usepackage{bm}

\newcommand{\bs}{\boldsymbol}
\def\be{\begin{equation}}
\def\ee{\end{equation}}
\def\beq{\begin{eqnarray}}
\def\eeq{\end{eqnarray}}
\def\bc{\begin{center}}
\def\ec{\end{center}} 

\issue{2020}{23}{4}{43702}
\doinumber{10.5488/CMP.23.43702}

\title[Heisenberg spin chains with additional]%
{Heisenberg spin chains with additional isotropic  
three-site exchange interactions}
\author[N.B. Ivanov]{N.B. Ivanov}
\address{
 Institute of Solid State Physics, Bulgarian 
Academy of Sciences, Tzarigradsko chauss\'{e}e 72, 1784 Sofia, Bulgaria
}
\date{Received June 15, 2020, in final form July 10, 2020}

\begin{document}
\maketitle
\begin{abstract}
The $J_1-J_3$ Heisenberg spin models  with nearest-neighbor ($J_1$) 
and additional isotropic three-site ($J_3$)  
spin interactions  remain relatively less explored, 
although such types of competing exchange terms can naturally emerge from 
different sources, including the strong-coupling expansion  of the 
multiorbital Hubbard model. Below we present a short survey of  
the recently published research in this field, the emphasis being  
on the  characteristics of the variety of quantum phases supported by 
a few  generic uniform-  and  alternating-spin $J_1-J_3$ Heisenberg chains.     
For the reason that  the positive ($J_3>0$) three-site 
couplings tend towards the formation of local quantum dimers,  
the $J_1-J_3$ spin models  typically experience  some spontaneous 
dimerization upon increasing $J_3$. Actually, it occurred  
that the established dimer phases in spin-$S$ $J_1-J_3$ Heisenberg chains 
($S>{1}/{2}$) serve as complete analogues of the famous gapped Majumdar-Ghosh 
dimer phase in the  spin-${1}/{2}$ Heisenberg chain with
next-nearest-neighbor couplings.
The same dimerizations have been observed in the  alternating-spin
($S,\sigma$) $J_1-J_3$ chains ($S>\sigma$), provided that the cell 
spin $S+\sigma=\rm{integer}$, whereas for half-integer cell spin, 
the local dimer formation produces gapless spin-liquid ground states. 
The alternating-spin  $J_1-J_3$ chains also  provide some typical examples 
of spin models supporting the  so-called non-Lieb-Mattis magnetic phases.
\end{abstract}

\section{Introduction}

Over the past two decades, it has been established that the Heisenberg spin
systems with additional competing interactions --- such as longer-range 
exchange bonds, Dzyaloshinskii-Moria couplings, as well as  ring and
biquadratic exchange couplings --- support a rich variety of spin phases, including  
the exotic spin ice and spin nematic states, as well as various spin 
liquids~\cite{frustration}.  In particular, the phase diagrams of Heisenberg  
spin models with the  two-site biquadratic exchange (2BE) term   
$\left(\bs{S}_i\cdot\bs{S}_j\right)^2$, where  $\bs{S}_i$ and 
$\bs{S}_j$ are lattice spins ($|\bs{S}_i|,|\bs{S}_j|>{1}/{2}$),  
were extensively  studied, typical examples  
being  the spin-1 bilinear-biquadratic  
chain~\cite{spin_1_chain} and its higher-dimensional counterparts on 
square~\cite{harada,spin_1_2D}, triangular~\cite{momoi,smerald}, 
and cubic~\cite{harada}  lattices.  

At the same time, by now the effect of the isotropic three-site 
exchange (3SE) terms   
\be\label{3SE}
\left( \bs{S}_i\cdot\bs{S}_j\right)\left(\bs{S}_i\cdot\bs{S}_k\right)+\mathrm{H.c.}
\hspace{1cm} (|\bs{S}_i|>1/2, \quad i\neq j, \quad k,j\neq k) 
\ee
on Heisenberg spin systems remains  relatively  less
explored. As a matter of fact, often both  higher-order exchange terms   
appear on an equal footing, for example,  in the fourth-order strong-coupling 
expansion of the two-orbital Hubbard 
model~\cite{bastardis}, so that finding higher-order exchange effects 
in real materials could  be a challenge. Indeed, to the best of our 
knowledge,  the only  more or less
convincing experimental evidence for 3SE effects comes from  inelastic 
neutron scattering results
for the  low-lying  excitations in the magnetic material CsMn$_x$Mg$_{1-x}$Br$_3$
($x=0.28$)~\cite{falk1},  CsMnBr$_3$ being  known as a nearly ideal 
isotropic one-dimensional (1D) Heisenberg antiferromagnet 
with site spins  $S={5}/{2}$.  
These experimental results  predict  almost identical strengths of both 
higher-oder exchange terms 
(2BE and 3SE),  which are about two orders of 
magnitude weaker than the principal bilinear Heisenberg coupling. 
It was established that the higher-order exchange interactions in   
CsMn$_x$Mg$_{1-x}$Br$_3$ also appear  as a result of magnetoelastic
forces~\cite{falk2}. Similar magnetostriction effects --- discussed earlier 
 for  polynuclear complexes of iron-group 
ions~\cite{iwashita} --- were predicted for 
some single-molecular magnets~\cite{furrer2}.    

In both mentioned scenarios for creating  higher-oder exchange
terms, 2BE and 3SE interactions are controlled by one and the same
parameter, so that it might  be  difficult to separate their effects 
in the experiment. Cold atoms in optical lattices open a promising route 
in this direction. It was demonstrated \cite{pachos1,pachos2,tame} 
that with the  two-species 
Bose-Hubbard model  in a triangular configuration, a wide range of Hamiltonian 
operators can  be generated, including different three-spin interactions. 
The latter are due to the possibility of atomic tunneling through 
different paths from one vertex to the other, and can be extended to  
1D spin models with three-spin interactions. 
Another intriguing system in optical lattices concerns  polar molecules  
driven by microwave fields, naturally giving  rise to Hubbard models with
strong nearest-neighbor three-site interactions~\cite{buchler}. Since the two-site
bilinear terms can be separately turned with external fields, this system 
opens  a promising route for experimental studies of such higher-order
exchange interactions. Let us also note that if one of the spins in the 2BE interaction 
is a spin-${1}/{2}$ operator, then this term reduces to an isotropic 
bilinear exchange. Thus, in some alternating-spin Heisenberg systems, 
the 2BE terms can be in principle excluded.

 On the theoretical side, generalized Heisenberg chains 
with additional 3SE terms have already been discussed in the 
literature, mostly as a tool to construct various  
isotropic~\cite{andrej,devega_woynar,aladim,devega,bytsko,ribeiro}
as well as  spin-${1}/{2}$ 
anisotropic~\cite{suzuki,gottlieb,titvinidze,lou,zvyagin,krokhmalskii2008,
derzhko2011,topilko2012,menchyshyn2015} exactly-solvable 
spin models. Note that the spin systems with anisotropic spin interactions go 
beyond the scope of the present survey, so that we restrict 
ourselves to a few comments.
As  mentioned above, in the case of spin-${1}/{2}$ operators,  
the expression in equation~(\ref{3SE}) reduces to a bilinear two-site term.
Therefore, the 3SE interaction should  necessarily contain products of three 
spin-${1}/{2}$ operators defined 
on separate lattice sites. Clearly, such additional  terms 
violate  the time and, eventually, the space reversal symmetries, so that they
might be expected to produce quite different effects as compared to the
isotropic case. Actually,  it was 
demonstrated that the discussed peculiarities 
of the 3SE terms in spin-${1}/{2}$ chains can  produce a 
number of intiguing effects, including the support  of 
some specific  phases such as the
chiral~\cite{gottlieb}, weak-ferromagnetic~\cite{zvyagin}, 
as well as   ferroelectric~\cite{menchyshyn2015} spin states.  

Turning again to the case of Heisenberg spin chains with 
isotropic 3SE interactions,  
note that only in  the recent few years the quantum phase diagrams  
of such models with arbitrary strengths ($J_3$) of the  3SE 
couplings were  discussed in the literature. 
Here, we present  a short  survey of the recent research in this field, 
the emphasis being on a few  generic 1D 
Heisenberg spin models with extra 3SE couplings, including
the spin-1 and spin-${3}/{2}$  uniform
chains~\cite{michaud1,michaud2,wang,chepiga1,chepiga2,chepiga3,chepiga4}, 
as well as the mixed-spin ($1,{1}/{2}$) and (${3}/{2},{1}/{2}$) 
chains~\cite{iv1,iv2,iv3,iv4}. The interest in such generalized $(J_1-J_3)$
isotropic spin models is partially motivated by the belief  
that the competing 3SE interactions
could  produce specific  phase diagrams, which are not typical of  
spin systems defined on frustrated lattices and/or with extra well-studied
competing interactions such as the longer-ranged exchange bonds and 
the 2BE interactions. Actually, as discussed below, the 3SE couplings exhibit
some unique features --- like the promotion of collinear classical 
spin configurations, or the reinforcing tendency towards clustering of 
the quantum spins on the shortest exchange bonds --- which may stabilize
some new spin phases.

The survey  is organized as follows. 
In section~\ref{uniform} we discuss recent works related to  
the quantum phase diagrams of two  spin-$S$ Heisenberg chains 
($S=1$ and ${3}/{2}$) and  extra isotropic 3SE
interactions. The same issues, but addressed to the 
alternating-spin Heisenberg chains with extra isotropic 3SE couplings and 
site spins $(S,\sigma)=(1,{1}/{2})$ and  
$({3}/{2},{1}/{2})$, are discussed in  section~\ref{mixed}. 
 Most of the presented  numerical results were obtained  using the
density-matrix renormalization group (DMRG) and exact numerical
diagonalization (ED) techniques. 
The last section contains conclusions and some prospects for future 
developments in the  field.  
\section{Spin-$S$ $J_1-J_3$ Heisenberg 
chains~\cite{michaud1,michaud2,wang,chepiga1,chepiga2,chepiga3,chepiga4}}
\label{uniform}
The Heisenberg spin-$S$ chain with additional isotropic 3SE interactions 
is defined by the  Hamiltonian
\be\label{h0}
{\cal H}_{1-3}= \sum_{i=1}^N \left\{ J_1 \bs{S}_{i}\!\cdot\!\bs{S}_{i+1}
+ J_3\left[\left(\bs{S}_{i-1}\!\cdot\!\bs{S}_{i}\right)
\left(\bs{S}_{i}\!\cdot\!\bs{S}_{i+1}\right)\!+\! \mathrm{H.c.}\right]\right\},
\ee
where $\bs{S_i}$ ($i=1,2,\ldots, N$) are site spins characterized by the
quantum spin number $S>{1}/{2}$, $N$ being the number of lattice sites. 
In the extreme quantum case
$S={1}/{2}$, the $J_3$ term reduces to a bilinear next-nearest-neighbor
(NNN) isotropic exchange interaction, as the squaring of any Pauli matrix gives the 
unit matrix. 
As a matter of fact, the most extreme quantum case with non-trivial $J_3$ terms
can be realized in spin systems with exchange bonds connecting only spin-1 and
spin-${1}/{2}$ operators, the alternating-spin models
discussed in the next section being perhaps the simplest and realistic examples of this kind. 
For convenience, we use the following parameterization  
$J_1 = J \cos \theta$ and $J_3 = J \sin \theta$, where 
$\theta \in [0, 2\piup ]$, and without loss of generality we set $J = 1$. 

\subsection{Spin-1 $J_1-J_3$  Heisenberg chain}
One of the first intriguing results, concerning  $J_1-J_3$ Heisenberg chains
with an arbitrary site spin $S$, states  that for some  strengths of 
the $J_3$ interaction, i.e., $J_1/J_3=[4S(S+1)-2]$,  the spin-S 
chain, equation~(\ref{h0}),  exhibits  an exact fully-dimerized (i.e., Majumdar-Ghosh type) 
ground state (GS)~\cite{michaud1}. This statement was soon generalized 
for $J_1-J_3$ Heisenberg chains with additional NNN
and alternating-bond interactions~\cite{wang}. 
Using the   conformal field-theory and DMRG numerical simulations, 
Chepiga et  al.  further analyzed  the spontaneous dimerization transition
in the spin-1 $J_1-J_3$ chain with additional NNN 
bonds~\cite{chepiga1,chepiga2,chepiga3}. 

\begin{figure}[ht]
	\includegraphics[width=0.48\textwidth]{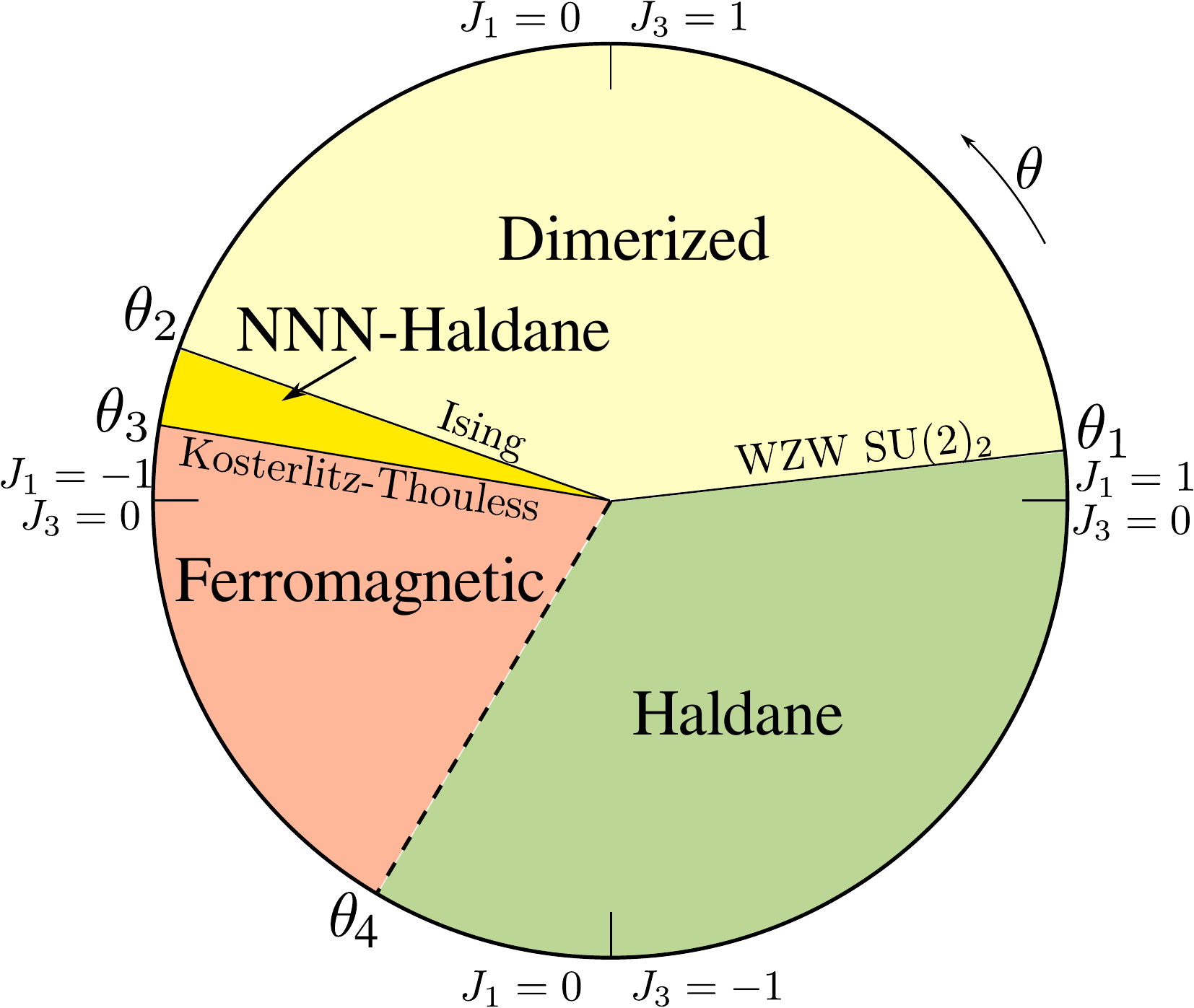}%
	\hfill%
	\includegraphics[width=0.48\textwidth]{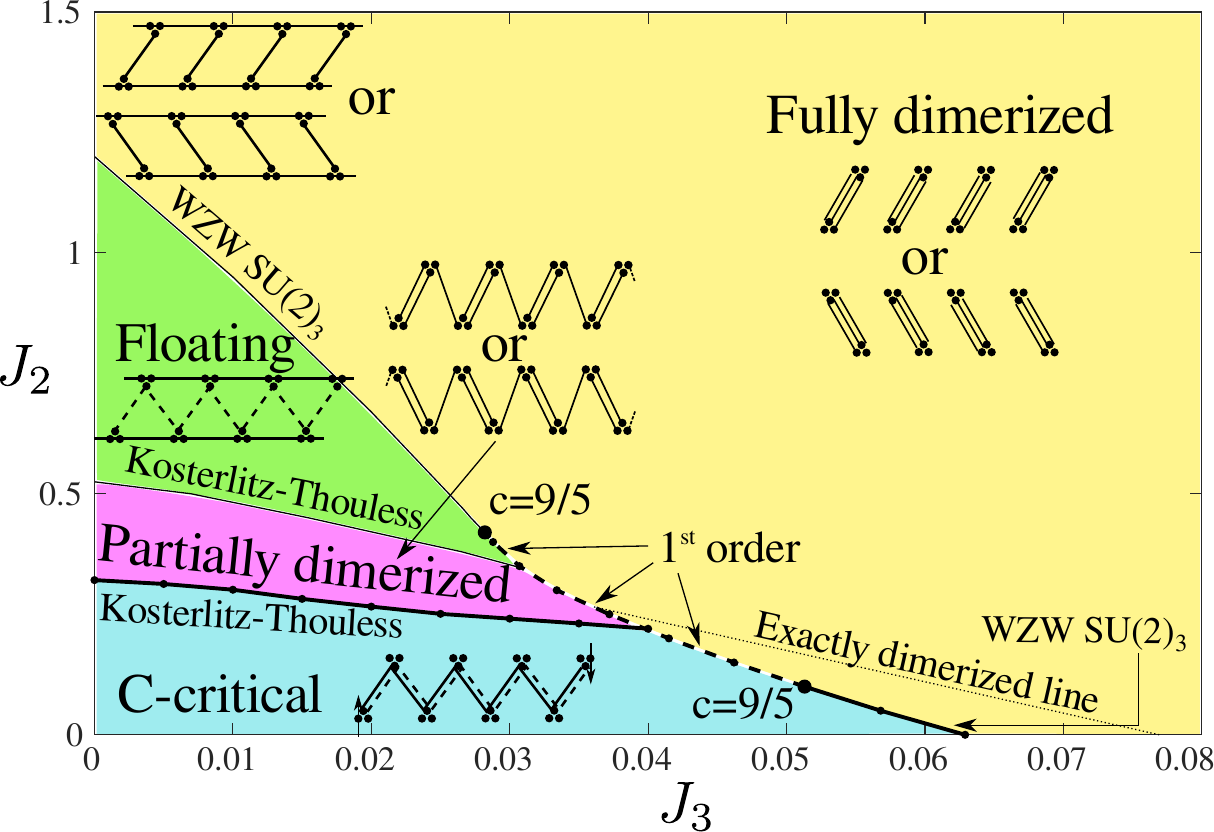}%
	\\%
	\parbox[t]{0.48\textwidth}{%
		\centerline{(a)}%
	}%
	\hfill%
	\parbox[t]{0.48\textwidth}{%
		\centerline{(b)}%
	}%
	\caption{(Colour online) (a) Phase diagram of the spin-1 Heisenberg chain with
		extra isotropic 3SE interaction ($J_1-J_3$  model) defined by equation~(\ref{h0}).
		$\theta_1\approx  0.03519\piup$, 
		$\theta_2\approx  0.8913\piup$,
		$\theta_3\approx  0.9474\piup$,
		$\theta_4\approx  1.33\piup$~\cite{chepiga2}.
		(b) Phase diagram of the spin-${3}/{2}$ Heisenberg chain  
		with next-nearest-neighbor ($J_2$) and 3SE ($J_3$) interactions. 
		The dimerized phases are gapped, whereas  the 
		C-critical and the floating phases are characterized by a gapless 
		spectrum. In terms of valence-bond states (VBS), the fully dimerized phase
		is represented as three valence bonds (solid lines) on every other $J_1$ bond, the 
		partially polarized phase is described by alternating one and two valence
		bonds, whereas the critical phase is visualized as one valence bond that
		resonates between two neighboring bonds (dashed lines)~\cite{chepiga4}.
	}
	\label{pd2}
\end{figure}

Turning to the quantum phase diagram of the spin-1 $J_1-J_3$ Heisenberg chain,
figure~\ref{pd2}~(a), we see that the spin system supports 
four different phases in the  interval $\theta \in [0,2\piup]$, i.e.,  Haldane,
dimerized, NNN-Haldane, and ferromagnetic (FM) phases.  
It was established  that the transition between the Haldane and dimerized phases at 
$\theta_1$  is continuous and belongs to the  Wess-Zumino-Witten  (WZW)
$SU(2)_{k}$ ($k=2$) universality class~\cite{michaud1}. Close to the pase boundary 
$\theta_1$ there is a dimerized point precisely at $J_3/J_1=1/6$. 

At $\theta_2$, an Ising-type transition takes place between the dimerized 
and the so-called NNN-Haldane phase~\cite{chepiga1,chepiga2} . 
The latter is a spin-1 counterpart of the 
dimerized Majumdar-Ghosh phase in the spin-${1}/{2}$ $J_1-J_2$ Heisenberg
chain --- $J_2$ being the strength of the NNN exchange bond --- but it consists 
of two effectively decoupled NNN Haldane chains~\cite{kolezhuk}.
   The DMRG data also suggest that in the thermodynamic limit the singlet-triplet
bulk gap remains open at $\theta_2$, while the spectrum becomes critical
within the singlet sector. The DMRG results  imply a direct 
Kosterlitz-Thouless phase transition at $\theta_3$ between the NNN-Haldane and the
FM phases. On the other hand, the transition between the FM phase and 
the Haldane phase at $\theta_4$ is first order~\cite{chepiga3}.
\subsection{Spin-${3}/{2}$ $J_1-J_3$  Heisenberg chains}
In the  spin-${3}/{2}$ $J_1-J_3$   chain, the transition 
from the critical phase to a spontaneously
dimerized state at $J_3/J_1\approx 0.063$~\cite{michaud2} is continuous and belongs
to the $SU(2)_k$ ($k=3$) WZW universality class~\cite{affleck1}. In
figure~\ref{pd2}~(b) we show the extended phase diagram of the 
spin-${3}/{2}$ $J_1-J_3$ Heisenberg chain with  additional NNN terms of
the form $\bs{S}_{i-1}\cdot\bs{S}_{i+1}$, which are controlled by the parameter
$J_2$~\cite{chepiga4}.

The phase diagram consists of two dimerized phases (partially and fully
dimerized) and  two critical phases with commensurate and
incommensurate correlations. Using the valence-bond singlets (VBS) picture, 
the fully dimerized phase corresponds to three valence bonds on every
other $J_1$ bond, while the partially dimerized phase corresponds
to alternating one and two valence bonds (see the insets in figure~\ref{pd2}).

In terms of VBS singlets, the commensurate C-critical phase can be
represented as one valence bond per $J_1$ bond, and, additionally,
one resonating bond between two neighboring bonds 
(the dashed lines). By contrast, the other critical phase --- called 
\textit{floating} phase --- appears for larger $J_2$. It is 
characterized by an incommensurate  wave vector $q$ which
changes within the phase area. Similar  phases with a varying $q$ are 
supported by the mixed-spin $J_1-J_3$ chains, as well (see the next section). 
Further interesting details concerning the phase diagram of the extended spin-
${3}/{2}$ $J_1-J_2-J_3$ Heisenberg model can be found in the original
work~\cite{chepiga4}.  
\section{Mixed-spin $J_1-J_3$ Heisenberg chains~\cite{iv1,iv2,iv3,iv4}}\label{mixed}
In view of the numerous experimentally  accessible quasi-1D  systems described 
by the mixed-spin  Heisenberg model (see, e.g., 
references~\cite{furrer2} and~\cite{landee}), it is instructive  to analyze the
following alternating-spin variant  of the spin-$S$ $J_1-J_3$ Heisenberg
chain discussed above~\cite{iv1} 
\be\label{h2}
{\cal H}_{1-3}'=  
\sum_{n=1}^L \left\{ J_1 \bs{S}_{2n}\!\cdot\!\left(
\bs{\sigma}_{2n-1}\!+\!\bs{\sigma}_{2n+1}\right)
+ J_3\left[\left(\bs{S}_{2n}\!\cdot\!\bs{\sigma}_{2n-1}\right)
\left( \bs{S}_{2n}\!\cdot\!\bs{\sigma}_{2n+1}\right)\!+\! \mathrm{H.c.}\right]\right\}.
\ee
 Here,  $L$ stands for the number of elementary cells, 
each containing two different spins ($S>\sigma$). We shall use again  the 
standard parameterization of the coupling constants 
$J_1=J\cos t$ and  $J_3=J\sin t$ ($0\leqslant t <2\piup$), where $J=1$. 

Since the effective strength of
the extra term is controlled by the parameter $S\sigma J_3$, one  
expects that  this interaction may play an important role especially
in  $(S,{1}/{2})$ chains and rings  with large $S$ spins ($S\gg
{1}/{2}$).  In  the extreme quantum case $(S,\sigma)=(1,{1}/{2})$,
it was demonstrated that the Hamiltonian ${\cal H}_{1-3}'$ 
reproduces --- up to irrelevant constants --- the  Hamiltonian of   isotropic  
spin-${1}/{2}$ diamond chain with  an additional ring exchange in the plaquettes
 in the Hilbert subspace  where the  pairs  of ``up'' and ``down'' plaquette spins 
form pure triplet states~\cite{iv5}.  
As mentioned above, the  alternating-spin systems  provide another 
realistic onset for separating the effects of the  higher-order exchange
interactions 2BE and 3SE. In particular, the Heisenberg chain with 
alternating $S$ and $\sigma={1}/{2}$ spins
($S>{1}/{2}$) provides  a simple  example of this kind. Indeed,
according to the operator identity $\left( \bs{S}_i\cdot\sigma_j\right)^2\equiv 
-\bs{S}_i\cdot\sigma_j/2+S(S+1)/4$, the  biquadratic terms
in this system reduce to bilinear  isotropic exchange terms. 
\begin{figure}
	\includegraphics[width=0.48\textwidth]{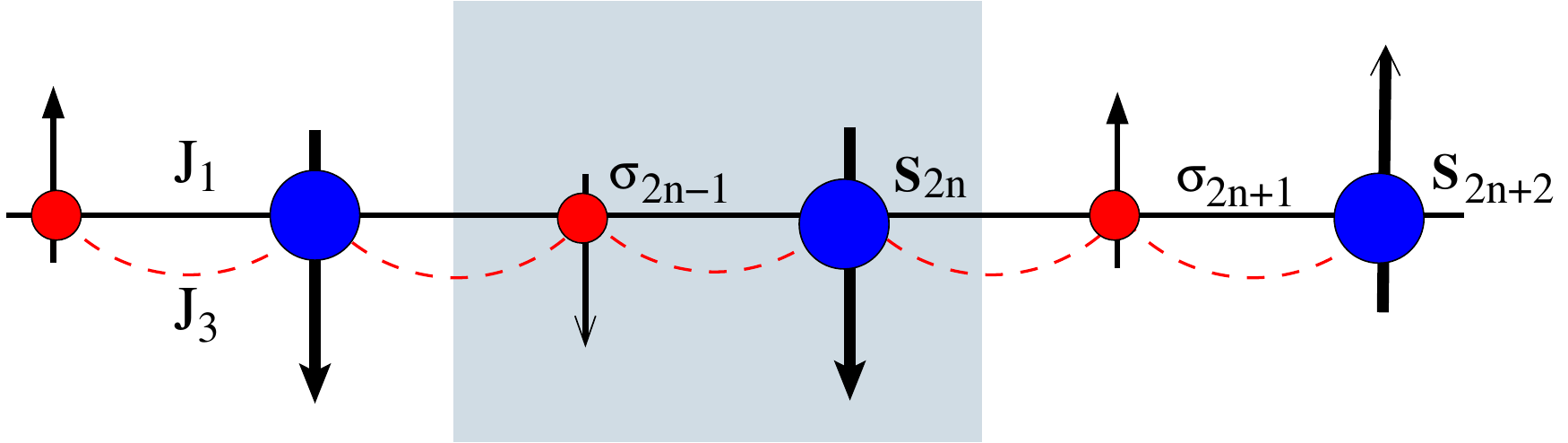}%
	\hfill%
	\includegraphics[width=0.48\textwidth]{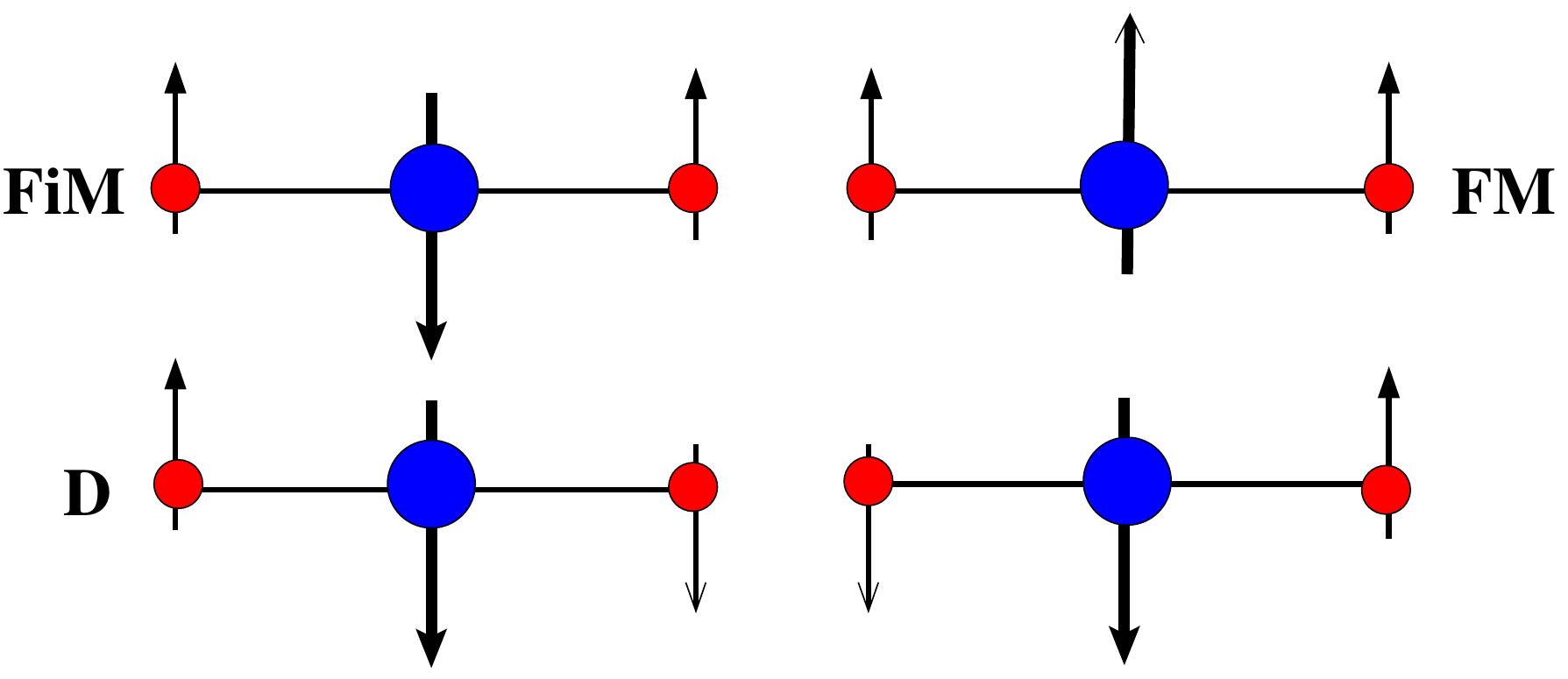}%
	\\%
	\parbox[t]{0.48\textwidth}{%
		\centerline{(a)}%
	}%
	\hfill%
	\parbox[t]{0.48\textwidth}{%
		\centerline{(b)}%
	}%
\caption{(Colour online) (a) Sketch of the mixed-spin  Heisenberg chains and (b) 
	the  optimal cluster states used as building blocks of the  classical 
	spin phases presented in figure~\ref{pd1}.
}
\label{chain}
\end{figure}

\begin{figure}[ht]
\centerline{	\includegraphics[width=0.9\textwidth]{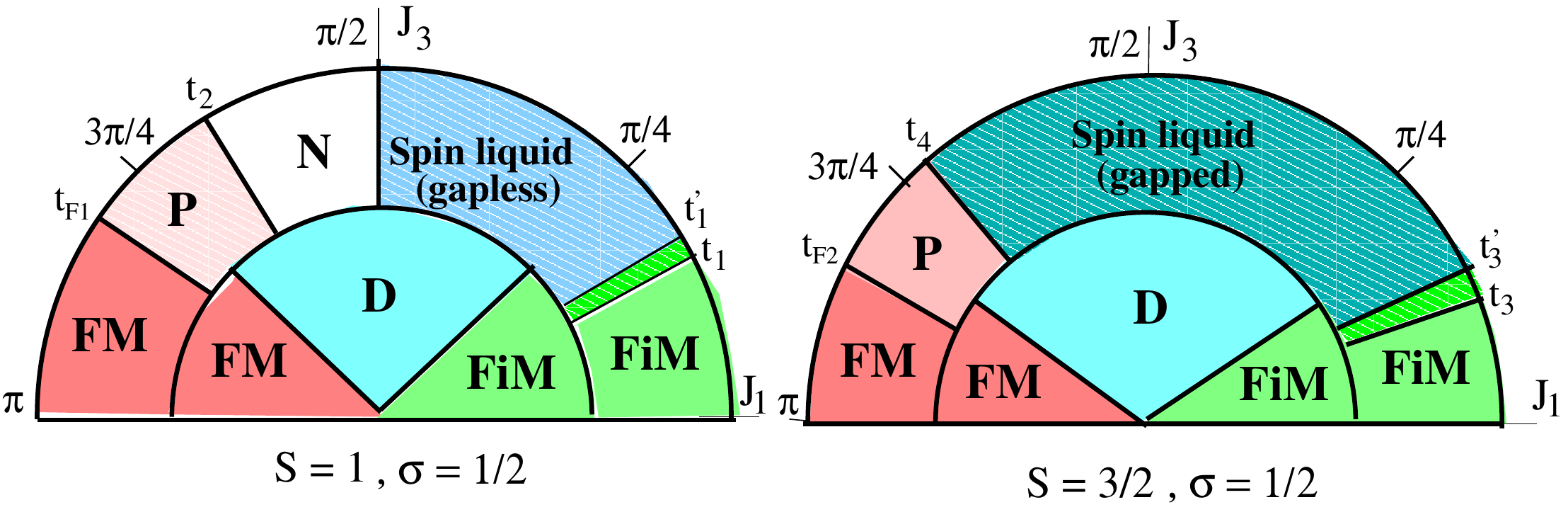}}%
\caption{(Colour online) 
	Classical (inner circles) and quantum  (outer circles) phase 
	diagrams of the $J_1-J_3$ mixed-spin chains with $(S,\sigma)=(1,{1}/{2})$
	(left-hand) and  $(S,\sigma)=({3}/{2},{1}/{2})$ (right-hand). 
	FM and FiM denote the collinear ferromagnetic
	and ferrimagnetic phases, respectively. $\bs{P}$ is a partially polarized 
	magnetic state with a spatially modulated magnetization. $\bs{D}$ denotes a
	$2^L$-fold degenerate classical phase composed of two 
	types of collinear cluster spin configurations shown in 
	figure~\ref{chain}. This phase contains both non-magnetic and magnetic 
	ground-state  configurations. The 
	gapless/gapped spin liquid  states correspond to chains with half-integer/integer
	cell spins $S+\sigma$. 
	$\bs{N}$ stands for a quasi-nematic state characterized by 
	short-range nematic order and spin-2  lowest-lying excitations. 
	Here, $t_1\approx 25.0^{\circ}$, $t_1'\approx 30.0^{\circ}$, $t_2\approx
	120.0^{\circ}$, $t_3\approx 20.1^{\circ}$, $t_3'\approx 25.5^{\circ}$, 
	and $t_4\approx 132.0^{\circ}$.
	$t_{\mathrm{F}}=\piup-\arctan [1/\sigma (2S+1)]$ is the exact FM phase
	boundary for arbitrary site spins $S$ and $\sigma$ ($t_{\mathrm{F}1}\approx
	146.3^{\circ}$, $t_{\mathrm{F}2}\approx
	153.4^{\circ}$). The classical phase boundaries between the FM and FiM
	phases of both models (not presented in this figure)  are 
	located at $t={3\piup}/{2}$. In the $(1,{1}/{2})$ case, quantum
	fluctuations stabilize a new doubly-degenerate (period two lattice cells)
	collinear magnetic  phase in the interval $253.1^{\circ}<t<264.0^{\circ}$, 
	whereas the boundary in the $({3}/{2},{1}/{2})$ case remains 
	unchanged~\cite{iv1,iv3}.   
}\label{pd1}
\end{figure}
\subsection{Classical phase diagram}
The classical phase diagram of the spin model~(\ref{h2}) 
(inner circles in  figure ~\ref{pd1}) 
can be constructed by using the GS 
cluster configurations shown in  figure~\ref{chain}. By fitting the
directions of the sharing $\sigma$ spins of the FM and FiM three-spin
clusters, one can obtain  macroscopic non-degenerate 
FM and FiM configurations corresponding to local minima of the classical 
energy, the FM and FiM phases in figure~\ref{pd1}. 
Using the two cluster configurations  denoted as $D$,
the same  procedure leads to a  $2^L$-fold  degenerate 
classical phase, denoted as $D$ phase in figure~\ref{pd1}. 
At a classical level,
the phase diagrams of both models are qualitatively indistinguishable.  

\subsection{Quantum phase diagrams}
\subsubsection{Modulated non-Lieb-Mattis type magnetic states}
In both models, the established partially-polarized  magnetic states ($P$ and
the narrow sectors close to the points $t_1$ and $t_3$ in figure~\ref{pd1})
do not appear  in the classical phase diagram. 
Due to the frustration effect of the three-site
interaction,  the Lieb-Mattis theorem~\cite{lieb} is not applicable, so that 
the so-called \textit{quantization} of the unit-cell magnetic moment is destroyed.
This means that the magnetic moment per cell can take arbitrary intermediate
values. In particular,  in such systems the magnetic moment may change 
continuously to zero near the transition from magnetic  to  non-magnetic
states ($t_1'$ and $t_3'$ points in figure~\ref{pd1}), or to be spatially  
modulated, like in the $P$ sectors of both models close to the
FM  phase boundaries $t_{\mathrm{F}1}$ and $t_{\mathrm{F}2}$. 
In the latter case,  the local magnetization 
clearly shows an incommensurate modulation with long-distance periodicity depending on the distance from the FM boundary. As a matter of fact, these
modulated partially-polarized states follow the Oshikawa-Yamanaka-Affleck rule 
$q(S+\sigma-m_0)=\textrm{integer}$~\cite{oshikawa}, where 
$q$ is the period of the modulated structure, and $m_0$ is the
magnetic moment per unit cell~\cite{iv1,iv3}. Extremely close to the
FM boundaries, $t\lesssim t_{\mathrm{F}1}$
($t\lesssim t_{\mathrm{F}2}$), DMRG results imply that the modulated magnetic 
structures are characterized by $(q,m_0)=(8,{9}/{8})$ and 
$(3,{5}/{3})$ for the $(1,{1}/{2})$ and $({3}/{2},{1}/{2})$
models, respectively.
\begin{figure}
	\includegraphics[width=0.38\textwidth]{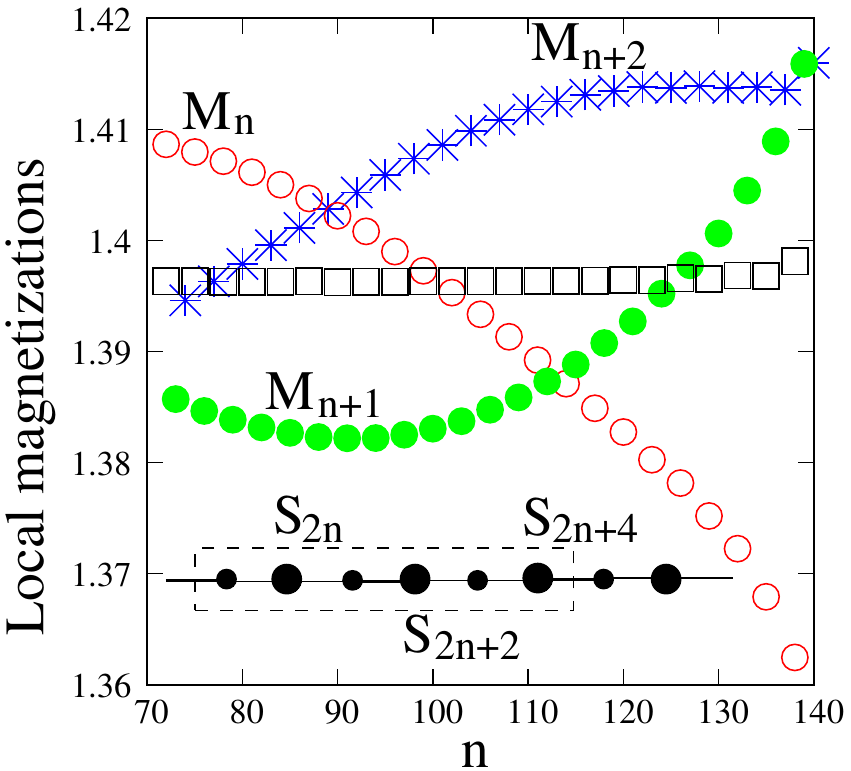}%
	\hfill%
	\includegraphics[width=0.58\textwidth]{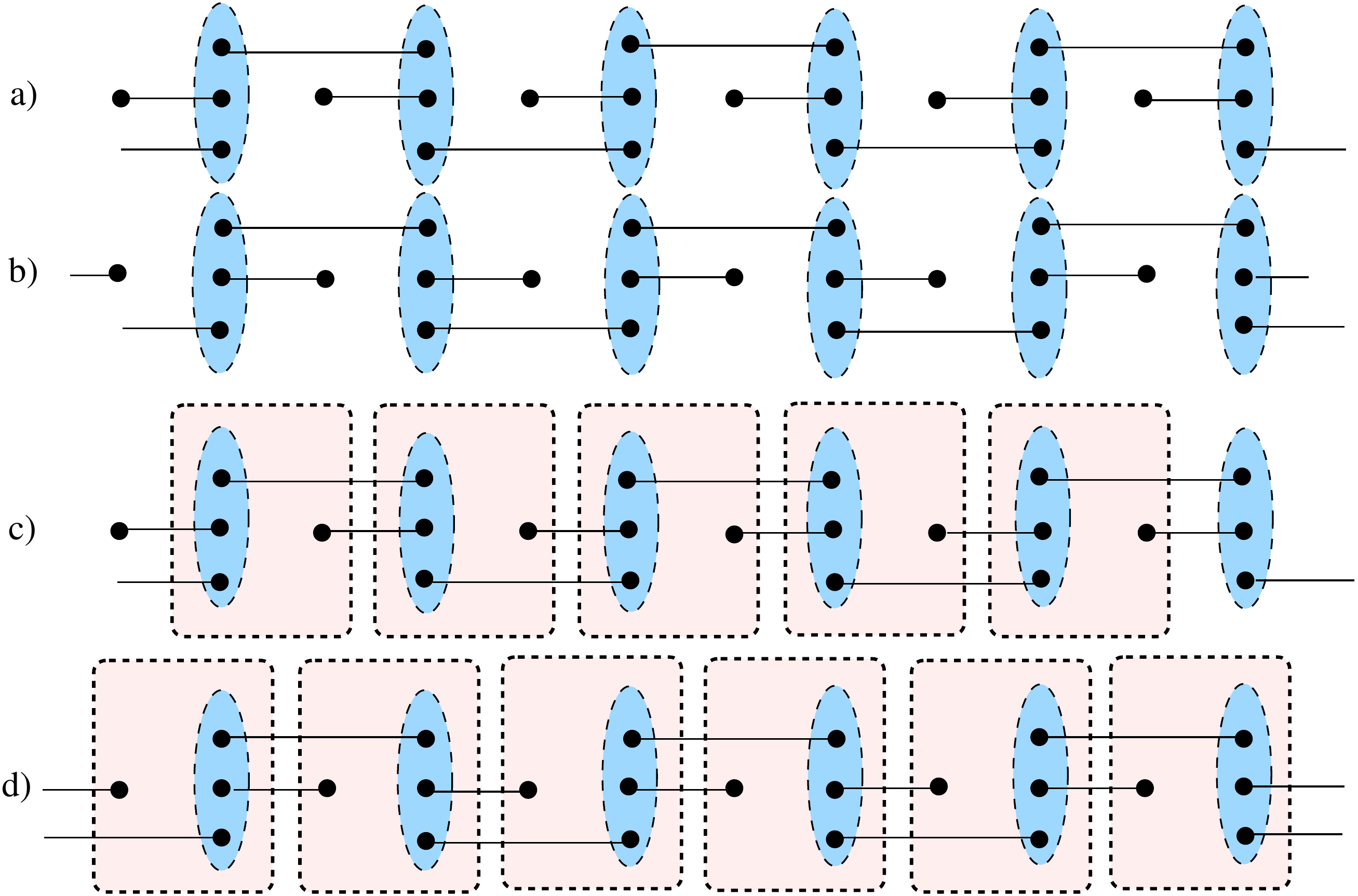}%
	\\%
	\parbox[t]{0.38\textwidth}{%
		\centerline{(a)}%
	}%
	\hfill%
	\parbox[t]{0.58\textwidth}{%
		\centerline{(b)}%
	}%
	\caption{(Colour online) 
		(a)  
		The on-site magnetizations $M_n=\langle S^z_{2n}\rangle$ in the P phase
		of the $({3}/{2},{1}/{2})$ chain  as functions of the cell index
		$n$ (DMRG, $t\approx 153.4^{\circ}$, $L=144$, open boundary conditions).  
		Here, $n=3j-2,3j-1,3j$ and  $j=1,2,\ldots, L/3$. Squares represent the
		constant average magnetization $(M_{n}+M_{n+1}+M_{n+2})/3$
		including  three neighboring unit cells. The results  
		demonstrate the establishment of a periodic three-cell ($q=3$) 
		magnetic structure  close to the FM 
		transition point $t_{\mathrm{F}2}$ in the P phase.  The Inset shows the magnetic 
		supercell containing six spins (i.e., three unit cells). 
		(b) 
		VBS  picture of the doubly degenerate spin-liquid phase of
		the  $({3}/{2},{1}/{2})$ chain  close to the phase 
		boundaries $t_3'$  
		(a,b) and $t_4$ (c,d), respectively.  Black dots
		denote spin-${1}/{2}$ variables. The lines between two spins ${1}/{2}$
		denote a singlet bond, whereas  the dashed ellipses and rectangles denote symmetrization of
		the spin-${1}/{2}$ variables. The first two (the last two) VBS 
		states approximately
		represent ground states of the open spin-1 (spin-2) AFM Heisenberg chain. In the
		intermediate region, only a part of the composite cell spins form spin-2
		states~\cite{iv3}.     
	} 
	\label{pp}\end{figure}

 As an example, in figure~\ref{pp}~(a) we present 
the on-site magnetizations of the established $q=3$ magnetic structure 
in the alternating-spin $({3}/{2},{1}/{2})$ $J_1-J_3$ chain 
extremely close to the exact FM boundary $t_{\mathrm{F}2}$, but out of the FM phase. 
As required for a plateau state, at this point 
the established magnetization per cell $m_0=5/3$ exactly fulfills 
the mentioned  general rule for period  $q=3$.
Interestingly, the effect of the open boundaries in this case simply 
reduces to some local redistribution  of the magnetic moment in the 
framework of the supercell, but the parameter 
$m_0$, characterizing the supercell as a whole,  remains  practically constant 
with the cell index $n$, excluding some narrow region near the system boundary.   
 Skipping the further  discussions on  these exotic 
modulated partially-polarized  states, we  mention that  such 
non-Lieb-Mattis-type magnetic phases in spin systems were 
originally identified in \cite{iv5},  
and later were studied in a number of other   
1D frustrated spin models~\cite{,shimokawa,furuya}.
\subsubsection{Degenerated spin-liquid phases}
DMRG analysis of the short-range correlations in open chains reveals the
regions in  the parameter space where the lowest-energy states in both
alternating-spin ($J_1-J_3$) models exhibit a regular alternating-bond 
structure characterized by different 
values  of the spin-spin correlators 
$\langle\bs{\sigma}_{2n-1}\cdot\bs{S}_{2n} \rangle =u$
and $\langle\bs{S}_{2n}\cdot\bs{\sigma}_{2n+1}\rangle =v$ ($u<v$).
The  $uv$ ($vu$) dimerized GS $|\Psi_{\mathrm{L}}\rangle$ ($|\Psi_{\mathrm{R}}\rangle$)
is stabilized in  open chains with a $\sigma$ spin on 
the left (right) end of the chain and corresponds
to a  $uv$ ($vu$) dimerization.
 The established $uv$ structure of the GS is strongly revealed 
in the middle of the phase-diagram regions occupied by non-magnetic 
states of both models,
where the  values of $u$ and $v$  indicate 
the formation of nearly  pure  spin-${1}/{2}$ and spin-$2$ (or spin-$3$) 
states of the nearest-neighbor spins in the systems with 
half-integer and integer cell spins $S+\sigma$,
respectively,  see figure~\ref{pp}~(b). 
As mentioned above, similar dimerization effects of the 3SE
interactions also appear in  the spin-$S$ $J_1-J_3$ Heisenberg chains, the
NNN-Haldane phase being such a typical example. 
Since a given $\bs{S}_{2n}$ spin can form such local dimer states in two
different ways --- including  the left-hand or the right-hand nearest-neighbor $\sigma$ spin --- all 
phases  in the non-magnetic regions are doubly degenerate, irrespective of the 
low-energy structure of the spectrum (i.e., gapped or gapless). 
The finite-size scaling of the lowest triplet
excitation in the alternating-spin ($1,{1}/{2}$) $J_1-J_3$
Heisenberg model suggests a gapless doubly-degenerate  non-magnetic 
state --- as may be expected from the  dimerized  structure of 
GS --- constructed from local spin-${1}/{2}$ dimers. 
On the other hand, 
the dimerized spin-liquid  state of the 
alternating-spin $({3}/{2},{1}/{2})$ $J_1-J_3$ Heisenberg chain 
suggests the formation of either spin-$1$
(smaller $t$), or spin-$2$ (larger $t$) local dimers, which leads to the
formation of a gapped doubly-degenerated spin-liquid state  
in the whole non-magnetic region of the phase diagram.
\subsubsection{Nematic-like phase}
The $J_3$ interaction term related to the $n$-th unit cell, equation
(\ref{3SE}), can be represented in  the following symmetric form
\be\label{n}
V^{(3)}_n=\frac{1}{2}\sum_{\alpha,\beta}\left(S_{2n}^{\alpha}S_{2n}^{\beta}+
S_{2n}^{\beta}S_{2n}^{\alpha}\right)
\left(\sigma_{2n-1}^{\alpha}\sigma_{2n+1}^{\beta}+
\sigma_{2n-1}^{\beta}\sigma_{2n+1}^{\alpha}\right), 
\ee
where $\alpha,\beta= x,y$, and $z$. 
The  two symmetric forms  in the parentheses can
be considered --- up to some normalization
factors --- as tensor order parameters of  on-site and bond spin nematic phases 
constructed from $S$ and  $\sigma$ spins, respectively. Taking into consideration this form of 
$V^{(3)}_n$ let us speculate  that for dominating  isotropic 3SE 
interactions nematic orders of different types could be stabilized. 
\begin{figure}[htb]
	\includegraphics[height=6cm]{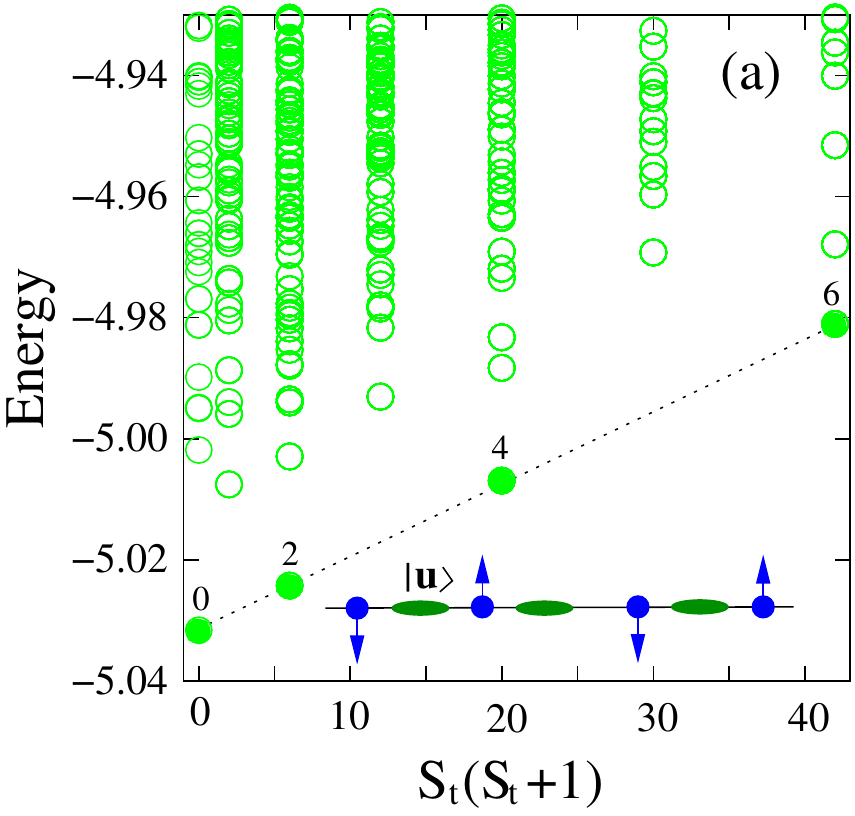}%
\hfill%
\includegraphics[height=6cm]{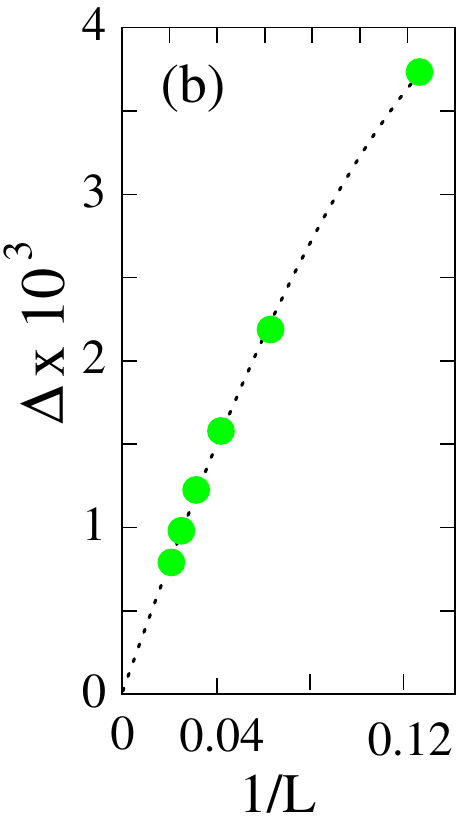}%
\hspace{25mm}
\\%
\parbox[t]{0.48\textwidth}{%
	\centerline{(a)}%
}%
\hfill%
\parbox[t]{0.48\textwidth}{%
	\centerline{(b)}%
}%
\caption{(Colour online) 
(a) Numerical ED results for the energy spectrum of the alternating-spin
($S=1$, $\sigma={1}/{2}$) ring ($L=8$, $t=110^{\circ}$)  
  \textit{vs} $S_{\mathrm{t}}(S_{\mathrm{t}}+1)$, where $S_{\mathrm{t}}$ is the total spin.  The lowest
 multiplets in the even $S_{\mathrm{t}}$ sectors (filled circles) form a 
tower of states with energies $E(S)\propto S_{\mathrm{t}}(S_{\mathrm{t}}+1)$.
The dashed line is a guide to the eye.  Inset: Cartoon of the suggested nematic-like 
state in the N sector constructed only with $S$ spins. 
The ellipses denote the local nematic states
$|\bs{u}\rangle=\sum_{\alpha}u^{\alpha}|\alpha\rangle$, where $\bs{u}$ is a
unit real vector and $|\alpha\rangle$ ($\alpha =x,y,z$) is the vector basis
of the spin-1 operator $\bs{S}$.
(b) Finite-size scaling of the gap to the lowest spin-2 excited state.
This state is the lowest excited state in the N region.  
The dashed line denotes the  least-squares fit to the DMRG data ($t=110^{\circ}$, OBC) 
obtained by the fitting ansatz $L \Delta(L)=a_0+a_1/L+a_2/L^2$~\cite{iv1,iv2}.   
} 
\label{dimer_tower}
\end{figure}
As a matter of fact,   the numerical ED calculations
concerning the lowest-lying excitations in periodic $L=8$ chains, described
by the Hamiltonian (\ref{h2}),  point towards the formation of
 some short-ranged nematic order only in the subsystem of $S$ spins~\cite{iv2},
 the $N$ sector in figure~\ref{pd1}. Unfortunately, DMRG  slowly
diverges, so that it is not effective, in this region of the phase diagram. 
However, some valuable
information on  the  non-magnetic $N$ state in the alternating-spin $(1,{1}/{2})$ 
$J_1-J_3$ Heisenberg model can also be extracted from the
lowest-lying states in different total-spin ($S_{\mathrm{t}}$) sectors. 
As well-known, the  established tower of well-separated 
 lowest  multiplets, containing only even $S_{\mathrm{t}}$
sectors, is a fingerprint of the spectrum of quadrupolar 
states (i.e., spin nematic phase), unlike  the Anderson tower of states~\cite{anderson} 
containing all $S_{\mathrm{t}}$ sectors. The latter is  a characteristic of the  
N\'{e}el  order. In fact, Anderson towers of states were observed  even in some 
finite  isotropic spin-$S$  chains and  magnetic molecules
\cite{schnack}. Interestingly, such regions 
with enhanced quadrupolar fluctuations do not appear on the phase diagram
of the $({3}/{2},{1}/{2})$ model.  Clearly, the discussed
non-magnetic phase deserves further studies.     

\section{Conclusion}
We discussed the  quantum phase diagrams 
of 1D isotropic spin models with extra  three-spin interactions,
the emphasis being on a few generic spin-1 and  alternating-spin ($J_1-J_3$) 
Heisenberg chains. As a whole, the models  are characterized  by rich  phase
diagrams, most of the phases being connected with some special features  of the
3SE interaction, equation (\ref{3SE}).  For example, a typical characteristic of the  discussed models is the presence of 
spontaneously dimerized phases in the upper part of the 
phase diagram ($J_3>0$). Clearly, this can be related to the fact that
for positive $J_3$ the 3SE terms promote the formation of local 
quantum dimers. 

Another interesting feature of the studied diagrams
is the appearance of some spatially modulated (non-Lieb-Mattis) 
partially-polarized  states between the collinear (FM and FiM) 
magnetic phases and the spin-liquid states in the alternating-spin models. 
Note that neither spiral nor spatially-modulated  magnetic intermediate states  
appear in  the classical phase diagram, which means that (i) the 
non-collinear spin configurations are suppressed by 
the interactions in equation (\ref{3SE}), and 
the appearance of intermediate states  represents a  pure quantum effect.

In conclusion, compared to some other widely discussed competing spin
interactions, the isotropic three-spin exchange reveals some unique
properties --- such as the promotion of  dimerization and linear spin
configurations --- which opens up an intriguing direction for future
theoretical and experimental studies in the field of quantum magnetism. 
On the theoretical side, studies of exotic spin phases 
in generic 1D and higher-dimensional spin models with extra 3SE 
couplings constitute an interesting and promising perspective. 
On the experimental side, it seems that cold atoms in optical lattices, 
as well as the large group of alternating-spin systems 
could  present  promising routes  to  construct different 
Heisenberg  spin systems with extra competing 3SE interactions. 
\section*{Acknowledgements}
I am grateful to Natalia Chepiga for the kind permission to
present her phase diagrams, and to J\"{u}rgen Schnack for  
collaboration in the  field.  This research was supported by the 
National Science Foundation of Bulgaria 
(Grant K${\Pi}$--06--X38/6/05.12.2019).

\ukrainianpart

\title{Гайзенберґові спінові ланцюжки з додатковими ізотропними триспіновими обмінними взаємодіями}
\author{Н.Б. Іванов }
\address{
 Інститут фізики твердого тіла, Болгарська академія наук, Царіградско шосе 72, Софія 1784, Болгарія
}

\makeukrtitle

\begin{abstract}
\tolerance=3000%
$J_1-J_3$ Гайзенберґові спінові моделі з взаємодією найближчих сусідів ($J_1$) та додатковою ізотропною триспіновою взаємодією ($J_3$) залишаються менш дослідженими, хоча такі типи конкуруючих обмінних членів можуть природно виникати з різних причин, включаючи розвинення сильного зв'язку мультиорбітальної моделі Габарда. Нижче ми представляємо короткий огляд нещодавніх досліджень у цій області з наголосом на характеристики розмаїття квантових фаз, які підтримуються декількома типовими одновимірними $J_1-J_3$ моделями Гайзенберґа з однорідними і змішаними спінами. Оскільки додатні ($J_3>0$) триспінові взаємодії ведуть до формування локальних квантових димерів, $J_1-J_3$ спінові моделі зазвичай зазнають певної спонтанної димеризації при збільшенні $J_3$. Справді, виявляється, що встановлені димерні фази у спін-$S$ $J_1-J_3$ Гайзенберґових ланцюжках ($S>\frac{1}{2}$) служать повними аналогіями відомої щілинної димерної фази Маджумдара-Ґоша у спін-$\frac{1}{2}$ Гайзенберґовому ланцюжку з взаємодією між наступними після найближчих сусідами. Такі ж димеризації виявлено у $J_1-J_3$ ланцюжках зі змішаними спінами-($S,\sigma$) за умови, що спін комірки $S+\sigma=\rm{integer}$, в той час як для пів-цілого спіна формування локальних димерів веде до безщілинних станів спінової рідини. $J_1-J_3$ ланцюжки зі змішаними спінами також передбачають деякі типові приклади спінових моделей, які реалізують так звані магнітні фази Ліба-Маттіса.
\keywords спінові ланцюжки, квантові спінові фази, триспінові взаємодії
\end{abstract}


\begin{thebibliography}{10}

\bibitem{frustration} Lacroix C., Mendels P.,  Mila F. (Eds.),
Introduction to Frustrated  Magnetism:
Materials, Experiments, \\Theory, 
Springer Series in Solid-State Sciences, Vol. 164, Springer, Berlin, Heidelberg, 2011,\\
\doi{10.1007/978-3-642-10589-0}.  

\bibitem{spin_1_chain} 
L\"{a}uchli A., Schmid G., Trebst S., 
Phys. Rev. B, 2006,  \textbf{74}, 144426,\\
\doi{10.1103/PhysRevB.74.144426}.

\bibitem{harada}
Harada K., Kawashima N.,
Phys. Rev. B,  2002, \textbf{65}, 052403,
\doi{10.1103/PhysRevB.65.052403}.

\bibitem{spin_1_2D} 
T\'{o}th T.A., L\"{a}uchli A.M., Mila F., Penc K.,
Phys. Rev. B,  2012, \textbf{85}, 140403(R),\\
\doi{10.1103/PhysRevB.85.140403}.

\bibitem{momoi}
Momoi T., Sindzingre P., Shannon N., 
Phys. Rev. Lett.,  2006, \textbf{97}, 257204,\\
\doi{10.1103/PhysRevLett.97.257204}.

\bibitem{smerald}
Smerald A.,  Shannon N., Phys. Rev. B,  2013, \textbf{88}, 184430,
\doi{10.1103/PhysRevB.88.184430}.  

\bibitem{bastardis} 
Bastardis R., Guih\'{e}ry N., de Graaf C.,
Phys. Rev. B,  2007, \textbf{76}, 132412,\\
\doi{10.1103/PhysRevB.76.132412}.

\bibitem{falk1}
Falk U., Furrer A., G\"{u}del H.U., Kjems J.K.,
Phys. Rev. Lett.,  1986,  \textbf{56}, 1956--1959,\\
\doi{10.1103/PhysRevLett.56.1956}.

\bibitem{falk2} 
Falk U., Furrer A., Kjems J.K., G\"{u}del H.U.,
Phys. Rev. Lett.,  1984, \textbf{52}, 1336--1339,\\
\doi{10.1103/PhysRevLett.52.1336}.

\bibitem{iwashita} 
Iwashita, T., Ury\^{u} N.,
J. Phys. Soc. Jpn.,   1974, \textbf{36}, 48--54,
\doi{10.1143/JPSJ.36.48}.


\bibitem{furrer2} 
Furrer A., Waldmann O.,
Rev. Mod. Phys.,  2013, \textbf{85}, 367--420, 
\doi{10.1103/RevModPhys.85.367}.


\bibitem{pachos1}
Pachos J.K., Plenio M.B.,
Phys. Rev. Lett.,  2004, \textbf{93}, 056402,
\doi{10.1103/PhysRevLett.93.056402}.


\bibitem{pachos2}
Pachos J.K., Rico E.,
Phys. Rev. B,  2004, \textbf{70}, 053620,
\doi{10.1103/PhysRevA.70.053620}.

\bibitem{tame}
Tame M.S., Paternostro M., Kim M.S., Vedral V.,
Phys. Rev. B,  2006, \textbf{73}, 022309,\\
\doi{10.1103/PhysRevA.73.022309}.

\bibitem{buchler}
B\"{u}chler H.P., Micheli A., Zoller P.,
Nat. Phys.,  2007, \textbf{3}, 726--731,
\doi{10.1038/nphys678}.

\bibitem{andrej}
Andrei N., Johannesson H.,
Phys. Lett. A,  1984, \textbf{100}, 108--112,
\doi{10.1016/0375-9601(84)90675-3}.

\bibitem{devega_woynar}
de Vega H.J., Woynarovich F.,
J. Phys. A: Math. Gen.,  1992, \textbf{25}, 4499--4516,
\doi{10.1088/0305-4470/25/17/012}.

\bibitem{aladim}
Aladim S.R., Martins M.J.,
J. Phys. A: Math. Gen.,  1993, \textbf{26}, L529--L534,
\doi{10.1088/0305-4470/26/12/009}.

\bibitem{devega}
de Vega H.J., Mezincescu L., Nepomechie R.I.,
Phys. Rev. B,  1994, \textbf{49}, 13223--13226,\\
\doi{10.1103/physrevb.49.13223}. 

\bibitem{bytsko}
Bytsko A., Doikou A.,
J. Phys. A: Math. Gen.,  2004, \textbf{37}, 4465--4492,
\doi{10.1088/0305-4470/37/16/001}.

\bibitem{ribeiro}
Ribeiro J.A.P., Kl\"{u}mper A.,
Nucl. Phys. B,   2008, \textbf{801}, 247--267,
\doi{10.1016/j.nuclphysb.2008.02.012}. 

\bibitem{suzuki}
Suzuki M.,
Prog. Theor. Phys.,  1971, \textbf{46}, 1337--1359, 
\doi{10.1143/PTP.46.1337}.

\bibitem{gottlieb}
Gottlieb D., R\"{o}ssler J.,
Phys. Rev. B,  1999, \textbf{60}, 9232--9235,
\doi{10.1103/PhysRevB.60.9232}.

\bibitem{titvinidze}
Titvinidze I., Japaridze G.I.,
Eur. Phys. J. B,  2003, \textbf{32}, 383--393,
\doi{10.1140/epjb/e2003-00113-8}. 

\bibitem{lou}
Lou P., Wu W.C., Chang M.C.,
Phys. Rev. B,  2004, \textbf{70}, 064405,
\doi{10.1103/PhysRevB.70.064405}.

\bibitem{zvyagin}
Zvyagin A.A., Phys. Rev. B,  2005, \textbf{72}, 064419,
\doi{10.1103/PhysRevB.72.064419}.

\bibitem{krokhmalskii2008}
Krokhmalskii T., Derzhko O., Stolze J., Verkholyak T.,
Phys. Rev. B,  2008, \textbf{77}, 174404,\\
\doi{10.1103/PhysRevB.77.174404}. 

\bibitem{derzhko2011}
Derzhko V., Derzhko O., Richter J.,
Phys. Rev. B,  2011, \textbf{83}, 174428,
\doi{10.1103/PhysRevB.83.174428}. 

\bibitem{topilko2012}
Topilko M., Krokhmalskii T., Derzhko O., Ohanyan V.,
Eur. Phys. J. B,  2012, \textbf{85}, 278, \\
\doi{10.1140/epjb/e2012-30359-8}.

\bibitem{menchyshyn2015}
Menchyshyn O., Ohanyan V., Verkholyak T., 
Krokhmalskii T., Derzhko O.,
Phys. Rev. B,  2015, \textbf{92}, \\184427,
\doi{10.1103/PhysRevB.92.184427}. 
\bibitem{michaud1} 
Michaud F., Vernay F., Manmana S.A., Mila F.,
Phys. Rev. Lett.,  2012, \textbf{108}, 127202,\\
\doi{10.1103/PhysRevLett.108.127202}.

\bibitem{michaud2} 
Michaud F., Manmana S.R., Mila F.,
Phys. Rev. B,  2013, \textbf{87}, 140404(R), \\
\doi{10.1103/PhysRevB.87.140404}. 


\bibitem{wang}
Wang Z.Y., Furuya S.C., Nakamura M., Komakura R.,
Phys. Rev. B,  2013, \textbf{88}, 224419,\\
\doi{10.1103/PhysRevB.88.224419}.

\bibitem{chepiga1}
Chepiga N., Affleck I., Mila F.,
Phys. Rev. B,  2016, \textbf{93}, 241108(R),
\doi{10.1103/PhysRevB.93.241108}.

\bibitem{chepiga2}
Chepiga N., Affleck I., Mila F.,
Phys. Rev. B,  2016, \textbf{94}, 205112(R),
\doi{10.1103/PhysRevB.94.205112}.

\bibitem{chepiga3}
Chepiga N., Mila F.,
Phys. Rev. B,  2019, \textbf{100}, 104426,
\doi{10.1103/PhysRevB.100.10442}.

\bibitem{chepiga4}
Chepiga N., Affleck I., Mila F.,
Phys. Rev. B,  2020, \textbf{101}, 174407,
\doi{10.1103/PhysRevB.101.174407}.

\bibitem{iv1}
Ivanov N.B., Ummethum J., Schnack J.,
Eur. Phys. J. B,  2014, \textbf{87}, 226,
\doi{10.1140/epjb/e2014-50423-7}.

\bibitem{iv2}
Ivanov N.B., Schnack J.,
J. Phys.: Conf. Ser.,  2014, \textbf{558}, 012015,
\doi{10.1088/1742-6596/1186/1/102014}.

\bibitem{iv3}
Ivanov N.B., Petrova S.I., Schnack J.,
Eur. Phys. J. B,  2016, \textbf{89}, 121,
\doi{10.1140/epjb/e2016-70057-y}.

\bibitem{iv4}
Ivanov N.B., Schnack J.,
J. Phys.: Conf. Ser.,  2019, \textbf{1186}, 012014,
\doi{10.1088/1742-6596/558/1/012015}.

\bibitem{kolezhuk}
Kolezhuk A.K., Schollw\"{o}ck U.,
Phys. Rev. Lett.,  1996,  \textbf{77}, 5142--5145, 
\doi{10.1103/PhysRevLett.77.5142}.

\bibitem{affleck1} 
Affleck I., Haldane F.D.M.,
Phys. Rev. B,  1987, \textbf{36}, 5291--5300, 
\doi{10.1103/PhysRevB.36.5291}.

\bibitem{landee}
Landee Ch.P., Turnbull M.M.,
Eur. J. Inorg. Chem., 2013, \textbf{2013}, 2266--2285,
\doi{10.1002/ejic.201300133}.

\bibitem{iv5}
Ivanov N.B., Richter J.,
Phys. Rev. B,  2004, \textbf{69}, 214420,
\doi{10.1103/PhysRevB.69.214420}.

\bibitem{lieb} 
Lieb E.H., Mattis D.C.,
J. Math. Phys.,  1962,  \textbf{3}, 749--751,
\doi{10.1063/1.1724276}.

\bibitem{oshikawa} 
Oshikawa M., Yamanaka M., Affleck I.,
Phys. Rev. Lett.,  1997, \textbf{78}, 1984--1987, \\
\doi{10.1103/PhysRevLett.78.1984}.

\bibitem{shimokawa}
Shimokawa T., Nakano H.,
J. Korean Phys. Soc.,  2013,  \textbf{63}, 591--595,
\doi{10.3938/jkps.63.591}.

\bibitem{furuya}
Furuya Sh.C., Giamarchi Th.,
Phys. Rev. B,  2014, \textbf{89}, 205131,
\doi{10.1103/PhysRevB.89.205131}. 

\bibitem{anderson}
Anderson P.W.,
Phys. Rev. B,  1952, \textbf{86}, 694--701,
\doi{10.1103/PhysRev.86.694}.

\bibitem{schnack}
Schnack J., Luban M.,
Phys. Rev. B,  2000, \textbf{63}, 014418,
\doi{10.1103/PhysRevB.63.014418}.

\end{thebibliography}
  \end{document}